\documentclass[doublespacing]{elsart}
\usepackage[english]{babel}

\usepackage{epsfig}
\usepackage{graphics}

\usepackage{amssymb}

\begin{document}

\begin{frontmatter}

\title{Superradiance Startup at Finite Temperatures of the Electromagnetic Field}

\author{S.V. Anishchenko}
\ead{sanishchenko@mail.ru}
 \address{Research Institute for Nuclear Problems, Bobruiskaya str., 11, 220030, Minsk, Belarus.}%
Corresponding author: S.V. Anishchenko, address  Research
Institute for Nuclear Problems, Bobruiskaya str., 11, 220030,
Minsk, Belarus.e-mail address: sanishchenko@mail.ru, sanishchenko@inp.bsu.by


\begin{abstract}
We use quantum-electrodynamical approach to study the initial  stage of Dicke superradiance from a system of two-level atoms. Applying the zeroth-order Magnus approximation, we
obtain the expression for the mean number of quanta emitted in the presence of
thermal fluctuations of the electromagnetic field.
\end{abstract}



\end{frontmatter}
\section{Introduction}

In 1954, Dicke  demonstrated that a system of $N_a$
inverted two-level atoms interacting with an electromagnetic field
can spontaneously drop to a ground state during the time
proportional to $N_a^{-1}$ \cite{Dicke1954}. This drop is
accompanied by the emission of an electromagnetic-radiation pulse
with the peak power $P$ proportional to $N_a^2$ and is called the
collective spontaneous emission, or  superradiance
\cite{Andreev1988}.

In the classical limit, Dicke derived a formula for time
dependence of the power $P$
 for two-level atom systems whose
dimensions are much less than the radiation wavelength. In
\cite{Eberly1971,Bonifacio1970,Bonifacio1971},
this formula was  generalized to the case of extended systems, but
the suggested theories -- except for the one given in
\cite{Bonifacio1970} and discussing single-mode
generation -- fall short to describe the initial stage of Dicke superradiant emission in
 a many-atom system.

Moreover, the question remains as to the effect of thermal
fluctuations of the electromagnetic field on superradiance. It is
well known \cite{Fain1958} that thermal fluctuations become
essential if $kT\ge\hbar\omega$; in fact, when $kT\gg\hbar\omega$,
generation starts as a stimulated emission induced by
thermal quanta rather than as a  spontaneous one.

In this context, our paper frames the quantum theory of
superradiance startup in the presence of thermal fluctuations of
the electromagnetic field. The paper is organized as follows.
First, with the zeroth-order Magnus approximation
\cite{Blanes2009,Blanes2010} we  find the expression for the
mean number  of photons, $N$, emitted by the system of
two-level atoms at the onset of the generation process  at finite
temperatures. We shall then demonstrate that under the
conditions of a single-mode generation   in the absence of
thermal fluctuations, the formula for $N$ converts to the
expression derived by Bonifacio and Preparata
\cite{Bonifacio1970}.

\section{Interaction of the electromagnetic field with two-level atoms }

Let us consider the emission processes involving a system of
two-level atoms. In the interaction representation, the
Hamiltonian of the system has the form
\begin{equation}
 \label{interaction1}
\hat H_{int}=\sum_\mu\hat a_\mu^+\hat b_\mu(t)+\hat a_\mu\hat b_\mu^+(t).
\end{equation}
Here $\hat a_\mu^+$ and $\hat a_\mu$ are the creation and
annihilation operators of  a photon with the energy
$\hbar\omega_\mu$ at the initial time. Generally noncommuting
operators  $\hat b_\mu^{(+)}$ depend on time and dynamical
operators characterizing the behavior of atoms. Using
~(\ref{interaction1}), we can easily write a formal expression for
the evolution operator $\hat U$ in the form suggested by Magnus
\cite{Blanes2009}
\begin{equation}
 \label{interaction2}
\hat U_{int}(t)=\exp\Big(-\frac{i}{\hbar}\int_0^t\hat H(x)dx-\frac{1}{2\hbar^2}\int_0^tdx\int_0^xdy[\hat H(x),\hat H(y)]+...\Big).
\end{equation}

Let us mention that the zeroth-order Magnus approximation was used
by Becker and  McIver \cite{Becker1987}, where they analyzed the photon
statistics in Cherenkov oscillators and free electron lasers (the Hamiltinian used in \cite{Becker1987} has the same form as in~(\ref{interaction1})):
\begin{eqnarray}
 \label{interaction3}
&\hat U_{int}\approx e^{-\frac{i}{\hbar}\int \hat H dt}=e^{\sum_\mu\hat a_\mu^+\hat \alpha_\mu(t)-\hat a_\mu\hat \alpha_\mu^+(t)},\nonumber\\
&\hat \alpha_\mu(t)=-\frac{i}{\hbar}\int\hat b_\mu(t)dt.\nonumber\\
\end{eqnarray}
The operators $\hat b_\mu^{(+)}(t)$ in \cite{Becker1987} were
assumed to be $c$-numbers. This only provided the
description of spontaneous radiation of relativistic electrons,
neglecting the recoil effect. Naturally, the statistics of the
emitted photons in this case is Poissonian, because $\hat U_{int}$
at $[\hat b_\mu^+(t)$, $\hat b_\mu(t)]=0$ corresponds to the shift
transformation responsible for a vacuum-to-Glauber state
transformation \cite{Glauber1963}.

Now, let us return to our problem. In the case under
consideration, the rate of energy exchange between the atoms and
the field is noticeably less than the characteristic oscillation
frequencies $\omega_\mu$ of the field. For this reason, the
application of the zeroth-order Magnus approximation seems
justified\footnote{The zeroth-order Magnus approximation describes exactly a system of two-level atoms resonantly
interacting with a single-mode radiation field, since for this
system the Hamiltonian in the interaction representation is
time-independent \cite{Andreev1988}.}. The fact that the operators
$\hat b_\mu^+(t)$ and $\hat b_\mu(t)$ are noncommuting is
essential, because it is just  for this noncommutativity that the
radiation is  enhanced, which we shall demonstrate  further in
this paper.

The time evolution of the annihilation operator ~$\hat a_\mu(t)$
in the zeroth-order Magnus approximation is given by
\begin{eqnarray}
 \label{interaction4}
&\hat a_\mu(t)=\hat U_{int}^+(t)\hat a_\mu(0)\hat U_{int}(t)\nonumber\\
&=\hat a_\mu(0)+[\hat A,\hat a_\mu(0)]+\frac{1}{2!}[\hat A,[\hat A,\hat a_\mu(0)]]+\frac{1}{3!}[\hat A,[\hat A,[\hat A,\hat a_\mu(0)]]]+...,\nonumber\\
&\hat A=-\hat a_\mu^+\hat \alpha_\mu(t)+\hat a_\mu\hat \alpha_\mu^+(t).\nonumber\\
\end{eqnarray}
We shall approximately consider the commutators
\begin{equation}
\label{interaction5}
\beta_{\mu\nu}(t)=[\hat \alpha_\mu^{+}(t),\hat \alpha_\nu(t)]\approx Tr\big([\hat \alpha_\mu^{+}(t),\hat \alpha_\nu(t)]\big)
\end{equation}
to be ordinary numbers, which is true when the quantum-mechanical
fluctuations of  $\beta_{\mu\nu}$ are negligibly small, and assume
 the commutators $[\hat \alpha_\mu^{(+)}(t),\hat
\alpha_\nu^{(+)}(t)]$ to be zero.
\begin{equation}
\label{interaction6}
[\hat \alpha_\mu^{(+)}(t),\hat \alpha_\nu^{(+)}(t)]\approx0.
\end{equation}
(As will be demonstrated further, the relationships
(\ref{interaction5}) and (\ref{interaction6}) strictly hold at the
superradiance startup.)

Then the operators ~$\hat a_\mu(t)$ readily transform to the form
 \begin{eqnarray}
 \label{interaction7}
&\hat a_\mu(t)=\sum_\nu\Big(\hat a_\nu\big(\cosh\sqrt{\beta(t)}\big)_{\nu\mu}+\hat\alpha_\nu(t)\big(\frac{\sinh\sqrt{\beta(t)}}{\sqrt{\beta(t)}}\big)_{\nu\mu}\Big)\nonumber\\
&=\sum_\nu \hat a_\nu C_{\nu\mu}+\hat\alpha_\nu S_{\nu\mu}.\nonumber\\
\end{eqnarray}

Assuming that the electric field at the initial time ($t=0$) was
at temperature $T$, we can find the mean number of quanta
in the mode with subscript~$\mu$ by formula
\begin{eqnarray}
 \label{interaction8}
&N_{\mu}(t)=<\hat a_\mu^+(t)\hat a_\mu(t)>=\sum_\nu\frac{1}{\exp(\hbar\omega_\nu/kT)-1}C_{\nu\mu}^*C_{\nu\mu}\nonumber\\
&+\sum_{\phi\nu}Tr(\rho_M\hat\alpha_\phi^+\hat\alpha_\nu)S_{\phi\mu}^*S_{\nu\mu}.\nonumber\\
\end{eqnarray}
A remarkable feature of (\ref{interaction8}) is that it needs
averaging  only over the variables characterizing the state of the
ensemble of elementary radiators, which can be done
using the density matrix $\hat\rho_M$ of the atomic subsystem.
Averaging over the field variables is already completed.

\section{Superradiance from two-level atoms}

By way of example, we shall consider in detail the startup of
superradiance from a system of two-level atoms resonantly
interacting with one field mode. In the interaction
representation, the Hamiltonian describing this resonant
interaction has the form \cite{Dicke1954,Andreev1988}
 \begin{equation}
 \label{superradiance1}
\hat H_{int}=-\kappa\hat a\hat R_{+}-\kappa^*\hat a^+\hat R_{-},
\end{equation}
where the atomic operators  $\hat R_+=\hat R_1+i\hat R_2$ and
$\hat R_-=\hat R_1-i\hat R_2$ satisfy the following commutation
relations \cite{Dicke1954,Andreev1988}:
\begin{eqnarray}
 \label{superradiance2}
&[\hat R_3,\hat R_\pm]=\pm\hat R_\pm,\nonumber\\
&[\hat R_+,\hat R_-]=2\hat R_3.\nonumber\\
\end{eqnarray}
Here $\hat R_1$, $\hat R_2$, and $\hat R_3$ are the three
projections of the pseudo-spin operator, each given by the sum of
Pauli operators
\begin{equation}
\label{superradiance3}
 \hat R_j=\sum_p\hat\sigma_j^{(p)},
\end{equation}
taken over all the atoms  ($j=1,2,3$).

The Hamiltonian  (\ref{superradiance1}) is associated with
 the following evolution operator
\begin{eqnarray}
 \label{superradiance4}
&\hat U_{int}=e^{\hat\alpha\hat a^+-\hat\alpha^+\hat a},\nonumber\\
&\hat\alpha=i\kappa^*\hat R_{-}t.\nonumber\\
\end{eqnarray}

Using the commutation relations  (\ref{superradiance2}), we find
the commutators needed for further calculations
\begin{eqnarray}
 \label{superradiance5}
&[\hat\alpha^+,\hat\alpha]=2\kappa^*\kappa\hat R_3t^2,\nonumber\\
&[\hat\alpha^{(+)},\hat\alpha^{(+)}]=0.\nonumber\\
\end{eqnarray}
The latter being equal to zero, the condition (\ref{interaction6})
is fulfilled.

Setting $\beta=2\kappa^*\kappa Tr(\hat R_3)t^2$ and using
(\ref{interaction8}), we obtain
\begin{eqnarray}
 \label{superradiance6}
&N=\frac{1}{\exp(\hbar\omega/kT)-1}|C|^2+\kappa^*\kappa t^2|S|^2Tr\big(\rho_M(\hat R^2-\hat R_3^2+\hat R_3)\big),\\
&C=\cosh(\sqrt{\beta}),\nonumber\\
&S=\frac{\sinh(\sqrt{\beta})}{\sqrt{\beta}}.\nonumber\\
\end{eqnarray}

The operator $\hat R^2=\hat R_+\hat R_-+\hat R_3^2-\hat R_3$
appearing in (\ref{superradiance6}) commutes with $\hat H_{int}$,
and so $\hat R^2$ is the integral of motion. In a system
containing $N_a$  inverted atoms, the
eigenvalues of operators $\hat R^2$ and $\hat R_3$ at the
superradiance startup equal $r(r+1)$ and $r=N_a/2$, respectively
  \cite{Dicke1954,Andreev1988}. Because the system is in a state
  corresponding to the eigenstate of the operator $\hat R_3$, the
  quantum-mechanical fluctuations of $\hat R_3$ vanish, thus
  supporting the validity of (\ref{interaction5}) for a system of
  two-level atoms.

Upon substituting  the eigenvalues of the operators $\hat R^2$ and
$\hat R_3$ into (\ref{superradiance6}), it  transforms to the form
\begin{equation}
 \label{superradiance4}
N=\frac{1}{\exp(\hbar\omega/kT)-1}\cosh^2(\sqrt{2\kappa^*\kappa r}t)+\sinh^2(\sqrt{2\kappa^*\kappa r}t).
\end{equation}
Let us note that at $kT\ll\hbar\omega$, the first term in
(\ref{superradiance4}) can be neglected, and
(\ref{superradiance4}) reduces to the expression derived in
\cite{Bonifacio1970}. As follows from Fig. 1, which  plots $N(t)$
against different radiation temperatures, the temperature
rise is accompanied by an increase in the  number of thermal
quanta at $t=0$.
\begin{figure}[ht]
\begin{center}
       \resizebox{75mm}{!}{\includegraphics{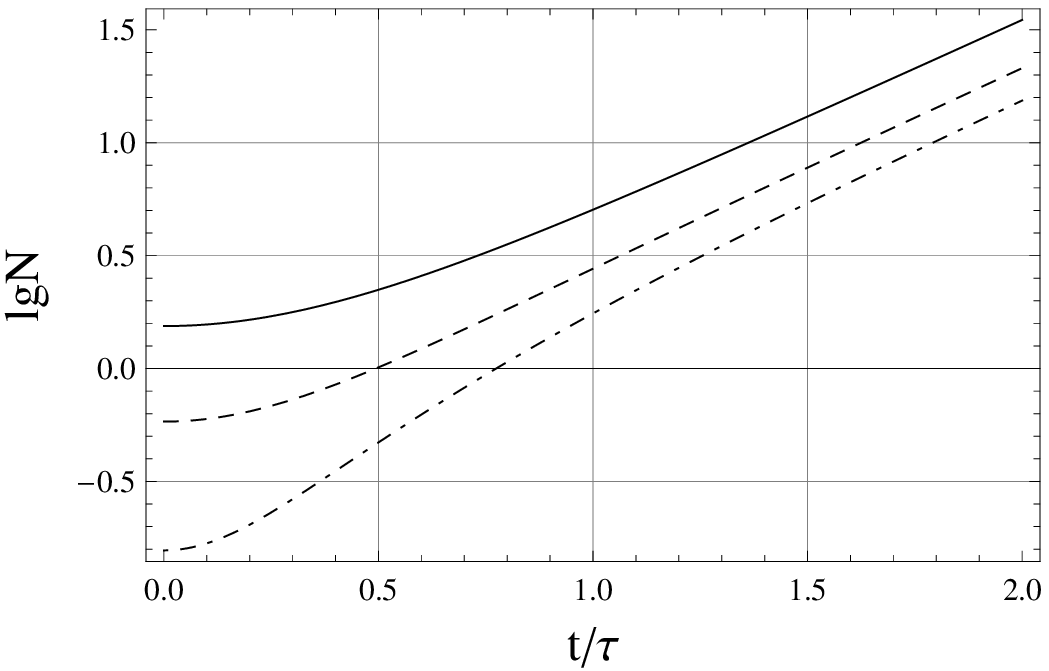}}\\
\caption{}
\end{center}
\end{figure}
When $kT\gg\hbar\omega$, generation begins as the emission induced
by thermal quanta rather than as a spontaneous emission.
As a result, the mean number of emitted quanta at the
superradiance startup is substantially increased (Fig. 1).

\section{Conclusion}
This paper frames the quantum theory of superradiance startup at
finite temperatures of the electromagnetic field. In the
zeroth-order Magnus approximation to the exponential perturbation
theory, we obtained the time dependence of the mean number of
emitted photons  at the generation startup. In the absence of
thermal fluctuations, this dependence coincides with that obtained
in \cite{Bonifacio1970}. In the reverse case, when $kT\gg \hbar
\omega$, the emission of photons becomes fundamentally different.
Actually, when $kT\gg\hbar\omega$,
 generation begins as the emission induced by thermal quanta instead of a spontaneous emission, and this fact should be taken into
account in spectroscopic studies and in development of terahertz
generators \cite{Siegel2002} operating at room temperature
($kT\sim\hbar\omega_{THz}$).

The suggested theory, based on the approximate relations
(\ref {interaction3}), (\ref{interaction5}), and
(\ref{interaction6}), is quite general in nature, because the
Hamiltonian (\ref{interaction1}) has the same structure as the
Hamiltonian describing the interaction of an arbitrary system of
charged particles with a transverse electromagnetic field. For
this reason,  the theory framed here applies to calculating the
radiation emitted by two-level atoms, as well as radiation emitted
by free charged particles in Cherenkov oscillators and free
electron lasers, which we shall demonstrate in subsequent  works.

\section{Acknowledgement}
I would like to thank Prof. V.G. Baryshevsky for posing a very
interesting question as to the necessity of quantum consideration
of the generation startup in many-particle systems.

Fig. 1. The number of photons as a function of time [solid curve:
 $\hbar\omega/kT=0.5$, dotted curve:  $\hbar\omega/kT=1.0$,
 and dash-dotted curve: $\hbar\omega/kT=2.0$,
$\tau=1/\sqrt{2\kappa^*\kappa r}$].

\end{document}